\documentclass[useAMS,usenatbib,referee]{mn2e}
\usepackage{amssymb}
\usepackage{graphicx}
\usepackage{threeparttablex}
\usepackage{longtable} 
\usepackage{fixltx2e}
\usepackage{lscape}
\title[Multi-wavelength Survey of Cluster AGN]{A Multi-wavelength Survey of AGN in Massive Clusters: AGN Distribution and Host Galaxy Properties}
\author[Alison J. Klesman and Vicki L. Sarajedini]{Alison J. Klesman$^{1}$\thanks{E-mail:
alichan@astro.ufl.edu (AJK); vicki@astro.ufl.edu (VLS)} and Vicki L. Sarajedini$^{1}$\\
$^{1}$University of Florida, Department of Astronomy, 211 Bryant Space Science Center, FL, 32611 USA}
\begin{document}

\maketitle

\label{firstpage}

\begin{abstract}
We investigate the effect of environment on the presence and fuelling of Active Galactic Nuclei (AGN) by identifying galaxies hosting AGN in massive galaxy clusters and the fields around them. We have identified AGN candidates via optical variability (178), X-ray emission (74), and mid-IR SEDs (64) in multi-wavelength surveys covering regions centered on 12 galaxy clusters at redshifts 0.5 $<$ z $<$ 0.9. In this paper, we present the radial distribution of AGN in clusters to examine how local environment affects the presence of an AGN and its host galaxy. While distributions vary from cluster to cluster, we find that the radial distribution of AGN generally differs from that of normal galaxies. X-ray-selected AGN candidates appear to be more centrally concentrated than normal galaxies in the inner 20\% of the virial radius, while becoming less centrally concentrated in the outer regions. Mid-IR-selected AGN are less centrally concentrated overall. Optical variables have a similar distribution to normal galaxies in the inner regions, then become somewhat less centrally concentrated farther from the cluster centre. The host galaxies of AGN reveal a different colour distribution than normal galaxies, with many AGN hosts displaying galaxy colours in the ``green valley" between the red sequence and blue star-forming normal galaxies. This result is similar to those found in field galaxy studies. The colour distribution of AGN hosts is more pronounced in disturbed clusters where minor mergers, galaxy harassment, and interactions with cluster substructure may continue to prompt star-formation in the hosts. Among normal galaxies, we find that galaxy colours become generally bluer with increasing cluster radius, as is expected. However, we find no relationship between host galaxy colour and cluster radius among AGN hosts, which may indicate that processes related to the accreting supermassive black hole have a greater impact on the star-forming properties of the host galaxy than does the intracluster medium and/or local galaxy environment. 

\end{abstract}


\section{Introduction}
Active Galactic Nuclei (AGN) are galaxies currently accreting significant amounts of material onto their central supermassive black holes (SMBHs). SMBHs are now believed to exist in the centres of all galaxies that contan a significant bulge component \citep{kr95}, and an observed relation between the mass of the SMBH and the bulge stellar mass \citep{fm00,geb00} suggests a coupled assembly history between the two. Galaxy clusters provide a unique environment in which to study the relationship between AGN and the galaxies in which they reside. 

The dense cluster environment is known to affect the evolution of cluster galaxies, producing a significant population of early-type galaxies in contrast to the lower-density field, which is comprised mainly of late-type galaxies \citep[e.g.,][]{hh31,morgan61,abell65,oemler74}. For AGN, the cluster environment could impact the onset and continued fuelling of the accretion process. The two most important fuel sources for AGN are believed to be cold gas reservoirs near the central black hole and galaxy mergers, which cause inflows of gas into the nucleus \citep[e.g.,][]{bh92,sdh05}. Since few cluster galaxies have significant reservoirs of cold gas \citep[e.g.,][]{gh85} and major mergers are rare in clusters due to high relative speeds between cluster galaxies, AGN were expected to be rare in clusters. The existence of an AGN would be an indicator that such a reservoir of available gas exists in the central region of the host galaxy. AGN can thus provide information about the efficiency of cold gas stripping in galaxy clusters, as well as the extent to which the central supermassive black hole may grow in such an environment. 

Cluster AGN can be used to probe galaxy/AGN evolution in dense environments \citep[e.g.,][]{galametz09,haines12}. Understanding the important factors in transforming star-forming galaxies into passive galaxies and the effects of environment on this process will provide a clearer picture of galaxy evolution. For example, it has been proposed that radiative feedback from the AGN may result in the quenching of star-formation and the transition to passively-evolving galaxies on the red sequence \citep[e.g.,][]{hopkins04, croton06, alexander05}. Feedback from AGN activity may also significantly contribute to the heating of the intracluster medium (ICM) and act as an important factor in galaxy evolution within the cluster \citep[e.g.,][]{mcnamara05}. Understanding the link between active galaxies and their environment is thus an important step in the process of understanding galaxy clusters and their evolution.

Early spectroscopic studies \citep{gisler78,dressler85} suggested that AGN were less common in rich clusters than in the field. However, more recent X-ray surveys have found an AGN fraction approximately five times higher than that of optical spectroscopic surveys \citep{martini02,martini06} and close to the X-ray AGN fraction in the field \citep{haggard10}. These results suggest that spectroscopic searches are missing many AGN, as emission from an AGN may be hidden due to weak line strengths, dust obscuration, or dilution of the source by the host galaxy light. A true understanding of the fraction of AGN in galaxy clusters thus relies upon the successful identification of AGN over a range of luminosities and obscuration levels.  

To address the AGN fraction in clusters and understand the role environment plays in their onset and continued fuelling, we studied a sample of 12 massive clusters at 0.5 $<$ z $<$ 0.9, in which AGN were identified via optical variability, X-ray emission and mid-IR power-law SEDs fitting. The identification of AGN candidates is presented in \citet{ks12a} (hereafter referred to as KS12). We identified an average of $\sim$25 AGN per cluster and found no significant difference between the fraction of AGN among galaxies in clusters and similarly-detected AGN in field galaxy studies ($\sim$2.5\%). This result indicates that the dense cluster environment does not appear to significantly hinder accretion onto the central SMBH to fuel an AGN.

In this paper, we examine the radial distribution and colours of AGN host galaxies in our clusters to investigate the impact of local environment on AGN and their hosts. In Section~\ref{clusters}, we briefly describe the galaxy cluster and AGN sample presented in KS12. In Section~\ref{radist}, we compare the radial distributions of AGN and normal galaxies in our cluster sample. In Sections~\ref{morph} and~\ref{AGNprop}, we describe how different cluster morphologies and AGN properties relating to obscuration may impact the AGN distribution in clusters. In Sections~\ref{colors} and~\ref{radcolors}, we discuss the overall distribution of galaxy colours among AGN as compared to normal galaxies in the cluster and as a function of cluster radius. Finally, we present our conclusions in Section~\ref{conclusions}. Throughout the paper we assume a standard cosmology with H$_{0}$ = 70, $\Omega_{M}$ = 0.3, and $\Omega_{\Lambda}$ = 0.7.

\section{Galaxy Cluster Sample and AGN Selection}
\label{clusters}

To investigate the role of environment on the AGN phenomenon, we conducted a census of AGN in several galaxy cluster fields (KS12). We chose clusters for which archival data were available that would allow AGN to be detected via optical variability, X-ray emission and mid-IR properties. With this multi-wavelength, multi-technique approach, we produced a largely unbiased sample with which to investigate the relationship between AGN, their environment, and their host galaxies. These same AGN identification techniques were also used to select AGN within the GOODS fields \citep{sarajedini11}, providing an excellent way to compare the field and cluster environments. All clusters were part of an HST ACS survey for supernovae in massive clusters \citep{sharon2010}. Every cluster had also been observed with the Chandra X-ray Observatory, while seven had mid-IR observations from the Spitzer Infrared Array Camera (IRAC). Table~\ref{clusterinfo} lists these clusters and several cluster properties.

\begin{table*}
\centering
\begin{minipage}{140mm}
\caption[Galaxy Cluster Properties]{Galaxy Cluster Properties \label{clusterinfo}}
\begin{tabular}{lcccr}
\hline
\multicolumn{1}{c}{Cluster} &
\multicolumn{1}{c}{Redshift} &
\multicolumn{1}{c}{M$_{200} (10^{15} M_{\odot})$} &
\multicolumn{1}{c}{r$_{v}$ (Mpc)} &
\multicolumn{1}{c}{Morphology Class\footnotemark[1]} \\
\hline
$CL0152-1357$          &  0.831   & 0.45  & 1.14 & Disturbed$^{b}$ \\
$CLJ1226.9+3332$    &  0.888    & 1.4   & 1.66  & Relaxed$^{b}$ \\
$MACSJ0257-2325$  &  0.506    & 1.41\footnotemark[2] & 	1.95 $\pm$ 0.42  & Semi-Relaxed$^{a}$ \\
$MACSJ0717+3745$ &  0.548    & 2.75\footnotemark[2] & 	1.93 $\pm$ 0.28  & Disturbed$^{a}$  \\
$MACSJ0744+3297$ &  0.686    & 0.87\footnotemark[2] & 	1.47 $\pm$ 0.18  & Semi-Relaxed$^{a}$  \\
$MACSJ0911+1746$ &  0.505    & 0.84\footnotemark[2] & 	1.61 $\pm$ 0.50  & Disturbed$^{a}$  \\
$MACSJ1149+2223$ &  0.544    & 1.13\footnotemark[2] & 	2.64 $\pm$ 0.14  & Disturbed$^{a}$  \\
$MACSJ1423+2404$ &  0.545    & 0.78\footnotemark[2] & 	1.35 $\pm$ 0.19  & Relaxed$^{a}$  \\
$MACSJ2214-1359$  &  0.504    & 1.22\footnotemark[2] & 	1.54 $\pm$ 0.16  & Semi-Relaxed$^{a}$  \\
$MS0451.6-0305$     &  0.550    & 1.4   &  2.6   & Relaxed$^{b}$ \\
$MS1054.4-0321$     &  0.830    & 1.1   &  1.8   & Semi-Disturbed$^{b}$ \\
$SDSS1004+41$       &  0.680    & 0.42 & 1.35  & Relaxed$^{b}$ \\
\hline
\end{tabular}
\footnotetext{$^{1}$ a: morphology classification from \citet{ebeling07}, b: morphology classification from \citet{ks12a}}
\footnotetext{$^{2}$ The value listed for M$_{200}$ corresponds to M$_{X,200}$ derived from X-ray gas as measured by \citet{barrett06}}
\end{minipage}
\end{table*}

We describe the AGN identification and selection process in detail in KS12, and thus provide only a brief summary here. Since our primary data set is the HST ACS images of each cluster, we limit our survey regions to the area slightly smaller than the ACS image centered at the cluster. This region is $\sim$3 $\arcmin$ across and corresponds to $\sim$1 Mpc at the cluster redshift. This is the region for which optical variability of the galaxies could be measured (i.e. where two or more ACS images of the cluster overlapped in multi-epoch imaging of the sources). Throughout the paper, we use the terms ``cluster regions'' or ``cluster field'' to refer to this optical variability survey field-of-view centered approximately at each cluster location.  
  
Each cluster in our sample had two to three epochs of HST ACS observations in the I band, with epochs separated by $\sim$one year. These data were used to photometrically identify varying nuclei in galaxies, which are likely to be AGN. We found 178 varying nuclei in $\sim$12,500 galaxies surveyed in the 12 cluster regions. These nuclei reach apparent magnitudes of I$_{nuc}$ $\leq$ 27, corresponding to absolute magnitudes of M$_I$ = -15.3 at z = 0.5 and M$_I$ = -16.8 at z = 0.9. We also identified AGN that appeared as X-ray point sources in each of our cluster regions using data from the Chandra Data Archive. These data were sensitive to point sources down to 7 x 10$^{-16}$ erg/cm$^{2}$/s in the full band. We found 74 X-ray point sources, with an average of six X-ray point sources per cluster region. Our X-ray flux limit in the full band corresponds to an X-ray luminosity of $\sim$6.7$\times$10$^{41}$ ergs/s at z = 0.5 and $\sim$2.8$\times$10$^{42}$ ergs/s at z = 0.9. Finally, AGN were identified in the infrared using Spitzer IRAC observations, which were available for seven of the 12 clusters. Power-law SED fits through the IRAC channels (3.6 to 8 microns) identify dust-obscured AGN via their re-radiated X-ray, UV, and optical light. We found 64 sources in our seven cluster regions with mid-IR power-law SEDs, resulting in $\sim$9 IR power-law sources per region. The limiting magnitude of the infrared survey was $\sim$18 in the 8-micron band. This corresponds to L$_{IR}$ = 1.4$\times$10$^{42}$ ergs/s at z = 0.5 and L$_{IR}$ = 6.0$\times$10$^{42}$ ergs/s at z = 0.9. Many of the AGN candidates were identified in more than one wavelength, with 24\% of X-ray sources and 4\% of mid-IR power-law sources also detected as optical variables. Combining all three detection techniques, we find an average of $\sim$25 AGN candidates per cluster region with a range of 12 to 49 AGN candidates in each region. 

\section{Radial Distribution of AGN}
\label{radist}

We compare the radial distribution of our AGN candidates to the radial distribution of normal galaxies for sources identified in our cluster fields. For this comparison, we define three samples based on the cluster membership probability of the galaxies and AGN. The first sample includes all sources found within the cluster survey region (i.e., everything within the overlapping ACS image field-of-view for the cluster). This sample will contain many foreground and background field sources, as well as cluster galaxies/AGN. The second sample contains those sources with $>$80\% probability of cluster membership as defined in KS12. These higher probability cluster members are generally those galaxies closer to the cluster centre, those with red colours, and spectroscopically-confirmed cluster members. The third sample contains only sources (AGN and galaxies) that are spectroscopically-confirmed cluster members and thus have 100\% probability of cluster membership.

\subsection{Individual Clusters}
We first compare the radial distribution of AGN to that of the normal galaxy population for our three samples in each individual cluster region. We perform a Kolmogorov-Smirnov test (KS test) to determine the probability that the AGN and galaxy radial distributions come from the same parent population. The results of these statistical tests are summarized in Table~\ref{rdistclusters}, along with other relevant cluster information. For each cluster, we give the redshift and fraction of the virial radius out to which we detect sources. We then report the KS probability that the radial distribution of AGN and normal galaxies are drawn from different parent populations for each of our three samples. We also give the total number of galaxies and AGN in the cluster field, as well as the number with greater than 80\% cluster membership probability and 100\% cluster membership probability (i.e., spectroscopically-confirmed cluster members).

\begin{table*}
\centering
\begin{minipage}{140mm}
\caption[Radial Distribution of Galaxies and AGN]{Radial Distribution of Galaxies and AGN \label{rdistclusters}}
\begin{tabular}{lcccc}
\hline
\multicolumn{1}{c}{Cluster} &
\multicolumn{1}{c}{Redshift} &
\multicolumn{1}{c}{KS\%$_{all}$ (\#Gal)(\#AGN)\footnotemark[1]} &
\multicolumn{1}{c}{KS\%$_{>80\%}$ (\#Gal)(\#AGN)\footnotemark[2]} &
\multicolumn{1}{c}{KS\%$_{100\%}$ (\#Gal)(\#AGN)\footnotemark[3]} \\
\hline
$CL0152-1357$    &  0.831 & 70.9 (727) (25) & 99.0 (185) (8) & 98.0 (39) (5)  \\
$CLJ1226.9+3332$ &  0.888 & 83.6 (937) (34) & 2.7 (581) (16) & 79.1 (25) (2)  \\
$MACSJ0257-2325$ &  0.506 & 87.6 (1188) (14) & 95.9 (1103) (14) & 78.3 (9) (1)  \\
$MACSJ0717+3745$ &  0.548 & 85.0 (1455) (46) & 52.1 (1141) (28) & 85.2 (81) (3) \\
$MACSJ0744+3297$ &  0.686 & 69.3 (1183) (30) & 70.8 (1090) (28) & 89.7 (27) (2) \\
$MACSJ0911+1746$ &  0.505 & 11.4 (1020) (19) & 42.2 (945) (16) & --- (15) (0)  \\
$MACSJ1149+2223$ &  0.544 & 15.5 (1444) (14) & 3.0 (1160) (11) & 76.7 (40) (1)  \\
$MACSJ1423+2404$ &  0.545 & 87.6 (1157) (51) & 86.6 (953) (37) & 79.2 (32) (7)  \\
$MACSJ2214-1359$ &  0.504 & 12.3 (1077) (10) & 54.7 (1020) (8) & 88.2 (41) (1)  \\
$MS0451.6-0305$  &  0.550 & 99.9 (834) (25) & 98.4 (334) (15) & 77.5 (55) (4) \\
$MS1054.4-0321$  &  0.830 & 99.5 (859) (26) & 77.8 (533) (11) & 96.6 (107) (4) \\
$SDSS1004+41$    &  0.680 & 69.3 (694) (20) & 73.8 (620) (18) & --- (0) (0) \\
All            &   --  & 99.6 (12575) (314) & 56.5 (9665) (210) & 99.3 (471) (30) \\
All (Variable AGN only) & -- & 52.0 (12575) (178) & 15.6 (9665) (132) & 89.1 (471) (30)  \\
All (X-ray AGN only) & -- & 79.3 (12575) (74) & 94.3 (9665) (47) & 87.8 (471) (11) \\
All (IR AGN only) & -- & 100.0 (12575) (64) & 62.5 (9665) (31) & 98.3 (471) (4) \\
Relaxed & -- & 94.9 (7070) (186) & 94.3 (5701) (136) & 95.2 (189) (17) \\
Disturbed & -- & 97.0 (5505) (131) & 43.9 (3964) (74) & 99.9 (282) (13) \\

\hline
\end{tabular}
\footnotetext{$^{1}$ The probability that AGN and galaxy radial distributions are drawn from different parent populations for all sources identified in the cluster field.}
\footnotetext{$^{2}$ The probability that AGN and galaxy radial distributions are drawn from different parent populations for sources with 80\% cluster membership probability.}
\footnotetext{$^{3}$ The probability that AGN and galaxy radial distributions are drawn from different parent populations for sources which are spectroscopically-confirmed cluster members.}
\end{minipage}
\end{table*}

Four of the twelve clusters in our survey show a significant difference between the AGN and normal galaxy distributions in at least one of our three samples. CL0152 shows significant differences among the samples with $>$80\% and 100\% probability of cluster membership, while MACSJ0257 shows a significant difference (96\% confidence) for the $>$80\% sample only. Examining the radial distributions of these two clusters reveals that the AGN sample is primarily less centrally concentrated than the normal galaxies except within the inner 20\% of the virial radius, where the AGN are either more centrally concentrated or closely follow the normal galaxy distribution. Two additional clusters, MS1054 and MS0145, reveal differences in the radial distribution of AGN and normal galaxies with even greater significance($<$97\% confidence for two of the three samples).

\begin{figure}
\centering
\includegraphics[scale=0.75]{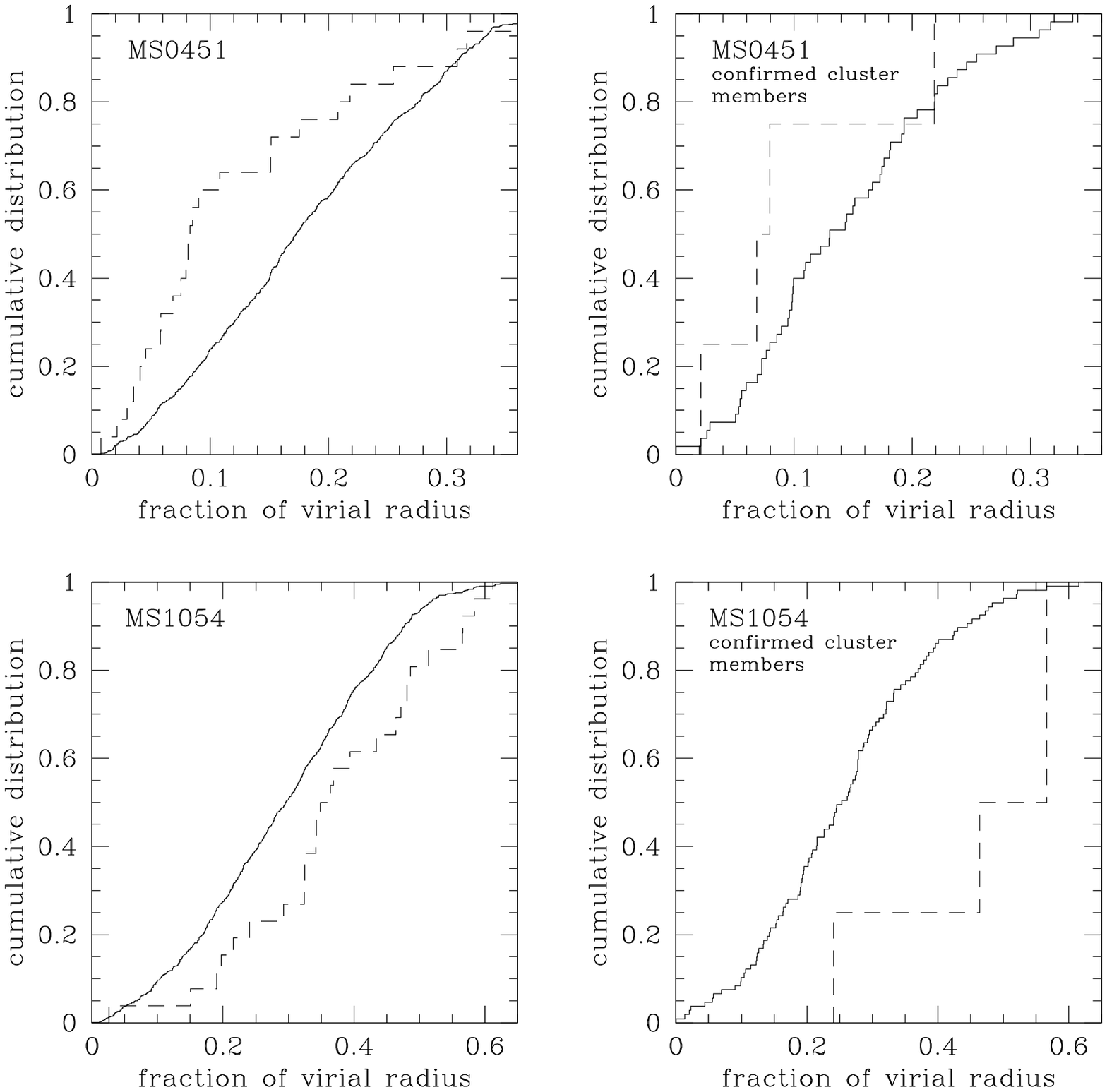}
\caption[Cumulative Distribution of Sources in 2 Cluster Fields]{Cumulative distribution of galaxies (solid line) and AGN (dashed line) as a function of virial radius fraction in MS0451 (top) and MS1054 (bottom) for everything within the cluster region (field and cluster sources) (left) and spectroscopically-confirmed cluster members only (right).}
\label{cumhistcomp}
\end{figure}

MS1054 and MS0145 are representative of the ways in which the radial distributions of galaxies and AGN differ from one another in our cluster sample. In Figure~\ref{cumhistcomp}, we show the cumulative distributions of galaxies in MS0451 (top panels) and MS1054 (bottom panels). The distribution for MS0145 extends to $\sim$32\% of the virial radius due to its low redshift (z = 0.55), whereas the distribution for MS1054 extends to more than 60\% of the virial radius at z = 0.83. What is most striking about these two clusters is the way in which their distributions differ from one another. In MS0145, the AGN are clearly more centrally concentrated than the normal galaxies, such that the differences are significant even in the sample containing many foreground and background sources (left, upper panel). In MS1054, we see the opposite trend. The AGN appear to be less centrally concentrated at all virial radii compared to the normal galaxies. To better understand these differences, we combine all cluster radial distributions in the next section and examine the distributions among different types of AGN. Based on our individual cluster analysis, we find that the AGN population is generally consistent with the distribution of cluster galaxies in two-thirds of our clusters, while one-third reveals AGN distributed differently from the normal galaxy population.

\subsection{Combined Cluster Distributions}

\begin{figure}
\centering
\includegraphics[scale=0.90]{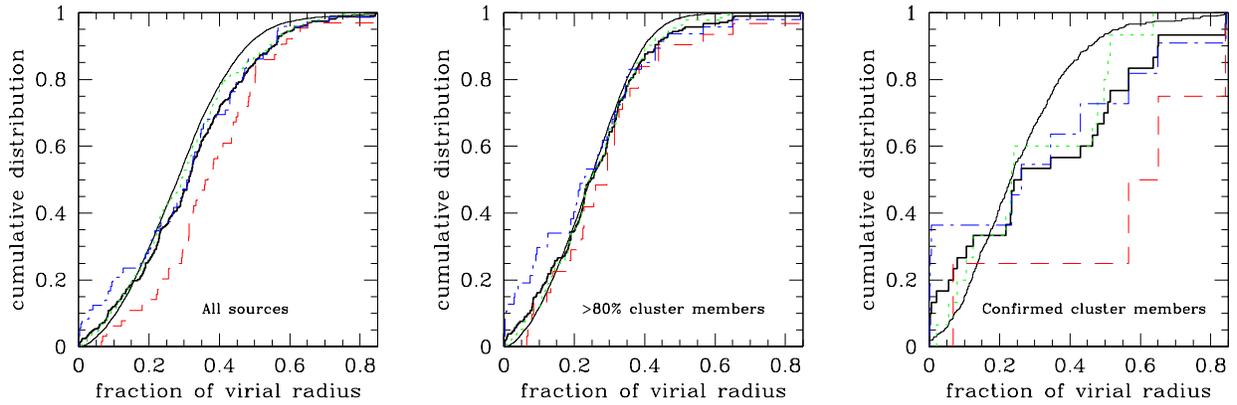}
\caption[Combined Cluster Cumulative Distribution]{Combined cumulative distribution of galaxies and AGN in the 12 survey cluster regions as a function of virial radius fraction for all galaxies and AGN in the cluster region (left), those with greater than 80\% probability cluster membership (middle), and spectroscopically-confirmed cluster members (right). The thick solid line represents the normal galaxies, the dashed line represents the AGN. Coloured lines represent the various types of AGN with optically varying AGN shown in green (dotted line), X-ray detected AGN shown in blue (dot-dashed line), and mid-IR power-law detected AGN shown in red (dashed line).}
\label{allch}
\end{figure}

In an effort to improve the statistics of our sample and explore differences with AGN type, we combine the radial distributions of all 12 clusters scaled by the virial radius. In Figure~\ref{allch}, the left panel shows all sources detected in the cluster fields. The central panel contains only those sources with $>$80\% cluster membership probability, while the right panel shows spectroscopically-confirmed cluster members only. We show the distribution for normal galaxies (solid line) and AGN (thick solid line), as well as the distribution for AGN detected using different identification techniques: optical variability (green dotted line), X-ray (blue dot-dashed line) and mid-IR SED power-law sources (red dashed line).

The KS test reveals significant differences in the radial distributions of AGN and normal galaxies among the first (all sources in the cluster field) and third (spectroscopically-confirmed only) samples ($>$99\% significance; see Table~\ref{rdistclusters}). The AGN radial distribution appears to be more centrally concentrated than normal galaxies within 20\% of the virial radius, then becomes less centrally concentrated than normal galaxies beyond this radius. The division of AGN types reveals that the X-ray sources (blue dot-dashed line) are responsible for the central concentration of AGN in the inner part of the clusters, while the IR sources (red dashed line) appear to be the least centrally concentrated of the different AGN types. The KS test for X-ray sources does indicate a significantly different distribution (94\%) among the sources with $>$80\% cluster membership probability and a slightly significant distribution difference (88\%) for the spectroscopically-confirmed cluster members only. IR AGN also reveal significantly different radial distributions (100\% and 98\% significance) among two of our three samples. The radial distribution of optical variables appears to lie somewhere between the two other types of AGN samples, revealing a distribution that follows the normal galaxies closely in the inner regions of the clusters, and then becomes slightly less centrally concentrated farther out with just under 90\% significance among the spectroscopically-confirmed cluster members.

Our result is consistent with \citet{galametz09}, who detected a slight (1.2$\sigma$) overdensity of X-ray sources in the centres of their clusters (r $<$ 0.5--1 Mpc) at redshifts similar to our clusters. Similarly, \citet{re05} find an excess of X-ray point sources within the inner r $<$ 0.5 Mpc of their sample of 51 MACS clusters. A recent study by \citet{ehlert13} finds that while their sample of clusters at similar redshifts contains an overdensity of X-ray sources, the X-ray sources are actually less centrally concentrated within the cluster than the normal galaxy distribution. This differs from our result, but may be explained with the following reasons: first, our X-ray survey contains fainter X-ray sources than the survey of \citet{ehlert13} (7$\times$10$^{-16}$ erg/cm$^{2}$/s in our survey, compared with 5$\times$10$^{-15}$ erg/cm$^{2}$/s; this corresponds to X-ray luminosities of 6.7$\times$10$^{41}$ erg/s in our survey and 4.8$\times$10$^{42}$ erg/s in their survey at z = 0.5). Second, we compare our AGN source distribution directly to that of the cluster galaxy distribution, rather than a King or NFW model distribution. If the difference is due primarily to the first reason, it may indicate that more luminous X-ray sources tend to avoid the centres of clusters, while less luminous ones prefer the denser regions, which is consistent with the results of \citet{kauffmann04}. Sensitive X-ray surveys of more clusters will provide additional data to test this theory.

\citet{galametz09}, also identified an overdensity of IR sources within the central 0.3 Mpc of clusters, which differs from our result. However, their IR sources were detected using mid-IR colours, rather than the power-law fitting approach that we use, and thus may contain a broader range of AGN candidates. \citet{atlee11} examine the radial distribution of X-ray point sources and IR power-law sources in a sample of low-redshift (z $\sim$ 0.06--0.31) galaxy clusters and find that both samples are consistent with the distribution of cluster galaxies, though with better agreement among the IR sources than the X-ray sources. We find an overall trend that the IR sources are less centrally concentrated than the normal cluster galaxies. 

\section{Cluster Morphology and Radial Distribution}
\label{morph} 
Galaxy clusters can have a range of morphologies and density distributions. In KS12, we discuss the morphology classifications for our clusters, which are fairly evenly divided between ``relaxed'' and ``disturbed" morphologies according to the criterion of \citet{re05}. We list these classifications in Table~\ref{clusterinfo}. \citet{re05} found that the observed central excess of X-ray sources among their clusters was more pronounced in relaxed clusters than disturbed ones. Relaxed clusters have a central cooling core dominated by their massive cD galaxies; \citet{re05} speculate that interactions (e.g., mergers, tidal interactions) with the cD galaxies and other giant ellipticals in the cores of these clusters could result in the excess of X-ray AGN they detect. 

\begin{figure}
\centering
\includegraphics[scale=0.90]{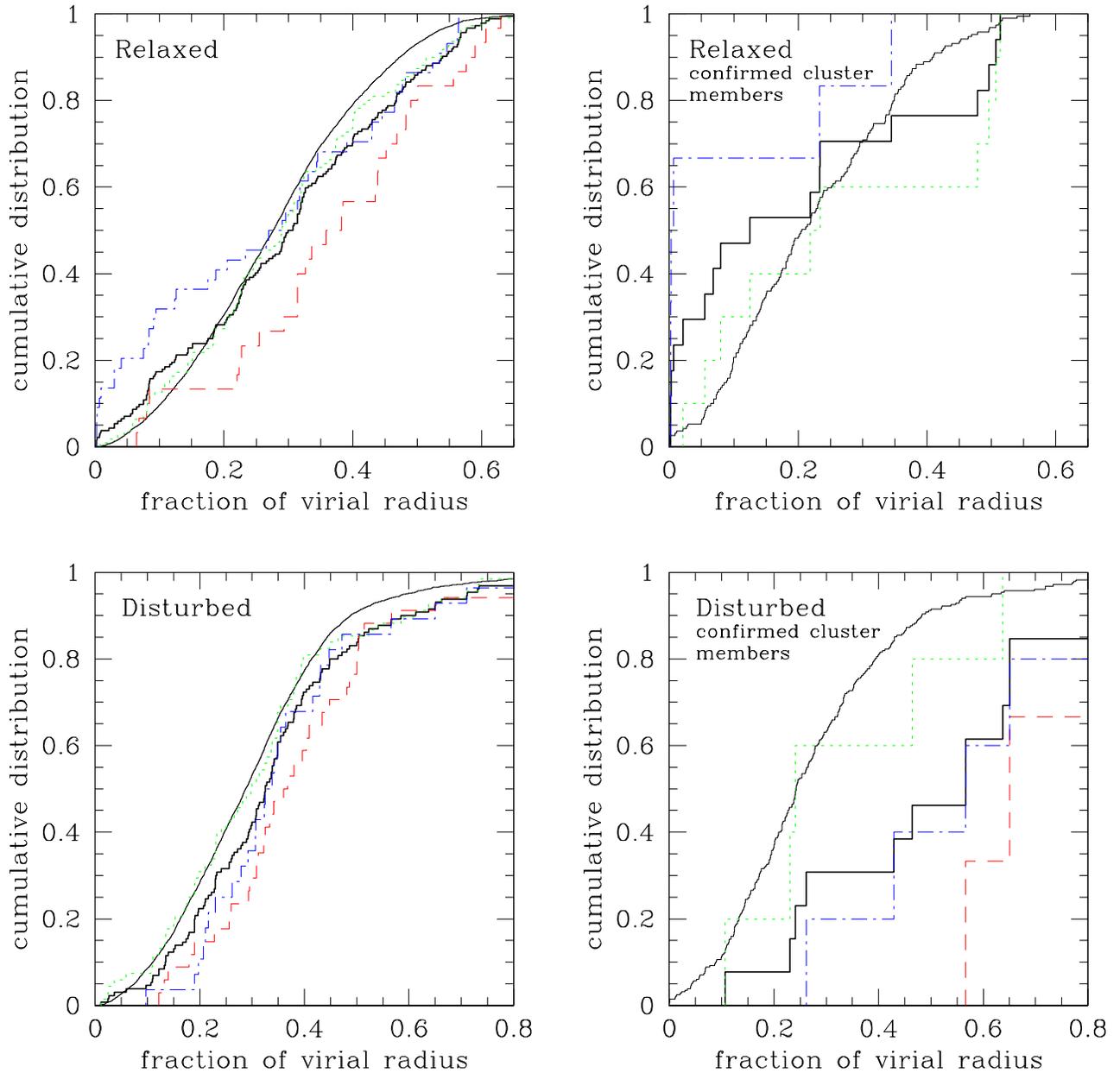}
\caption[Cumulative Distributions for Relaxed and Disturbed Clusters]{Cumulative distribution of galaxies and AGN in relaxed (upper panels) and disturbed (lower panels) cluster fields. Left panels show the distribution of galaxies (solid line) and AGN (thick solid line) for all sources in the cluster fields. Right panels show the same distributions for spectroscopically-confirmed cluster members. Coloured lines represent the various types of AGN with optically varying AGN shown in green (dotted line), X-ray-detected AGN shown in blue (dot-dashed line), and mid-IR power-law-detected AGN shown in red (dashed line).}
\label{agnchmorph}
\end{figure}

As shown previously, we find AGN distributions significantly different from the normal galaxy distributions in four of our 12 cluster regions. Of these four clusters, two are ``relaxed'' or ``semi-relaxed'' and two are ``disturbed'' or ``semi-disturbed.'' To further investigate the effects of cluster morphology on the AGN population, we compare the radial distributions of AGN and normal galaxies for relaxed and disturbed clusters in Figure~\ref{agnchmorph}. We combine the ``relaxed'' and ``semi-relaxed''  clusters (seven total) and the ``disturbed'' and ``semi-disturbed'' clusters (5 total) in this figure. In relaxed clusters, we find that the AGN population is more centrally concentrated than the normal galaxies for all sources in the cluster field (left, upper panel) and for spectroscopically-confirmed cluster members only (right, upper panel) at a 95\% significance level. The AGN appear more centrally concentrated within the inner $<$20--30\% of the virial radius, with the distribution dominated by the X-ray selected AGN (blue dot-dashed line). In contrast, the disturbed clusters do not have a centrally concentrated AGN distribution (lower panels). While the AGN distribution is still significantly different from the galaxy distribution (97\% significance and 99.9\% significance for the samples with all sources and spectroscopically-confirmed only, respectively), the AGN appear less centrally concentrated for all AGN types. Disturbed clusters by nature have significant substructure and their centres are not as well-defined. \citet{re05} explain that the lack of an AGN excess in the central regions of these clusters may be due to this fact, since any change in the distribution of AGN with respect to the cluster galaxies will be spread out over larger radii. 

\citet{re05} find an ``AGN depletion zone'' in relaxed clusters with a reduced density of AGN at radii $\sim$0.5 to 2 Mpc, followed by an increase of AGN at roughly the cluster virial radius where the cluster merges with the field. They explain this finding as a result of the shorter timescale for depleting a merger-induced accretion disk around a central supermassive black hole when compared with the cluster crossing time. \citet{re05} also suggest that the increase of AGN at the virial radius is the result of AGN triggered by mergers or galaxy harassment as galaxies first fall into the cluster. Our radial distribution results for relaxed clusters are not inconsistent with a depletion zone at intermediate cluster radii. The upper panels of Figure~\ref{agnchmorph} reveal an AGN distribution that rises much more slowly beyond the inner 30\% of the virial radius. This change in slope of the AGN distribution corresponds to a radius of $\sim$0.5 Mpc, since the average relaxed cluster virial radius is 1.7 Mpc. To quantify this, we calculate the ratio of AGN to normal galaxy number density as a function of virial radius. Within 0.1 virial radii for confirmed cluster members, we find the fraction of AGN to be 0.205$\pm$0.080 (8/39). This fraction drops to 0.049$\pm$0.019 (7/144) at virial radii between 0.1 and 0.5. From 0.5 to 0.6 virial radii, the fraction rises again to 0.333$\pm$0.272 (2/6). The drop between the inner and intermediate radius fractions is significant at just 1.6$\sigma$, and the rise beyond 0.5 virial radii is even less statistically significant due to the small number of sources. Therefore, these results appear largely consistent with the findings of \citet{re05}, though we cannot confirm their observed increase in AGN at the cluster virial radius.

\section{AGN Properties and Radial Distribution}
\label{AGNprop}

\begin{figure}
\centering
\includegraphics[scale=0.90]{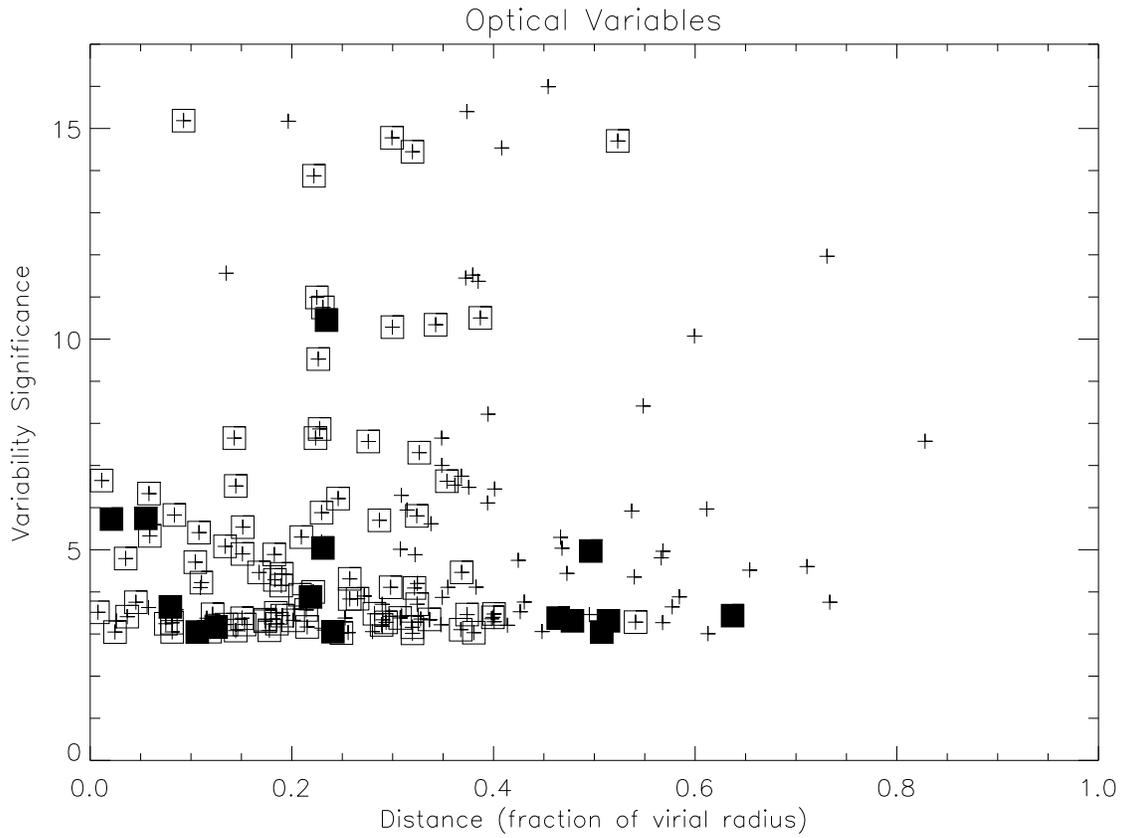}
\caption[Optical Variability Significance vs. Cluster Radius]{Variability significance vs. radial distance from the cluster centre. A plus symbol indicates all AGN in the cluster region, open squares indicate sources with $>$80\% cluster membership probability, and filled squares indicate spectroscopically-confirmed cluster members.}
\label{sigmavsdist}
\end{figure}

Next we explore correlations between the radial location of an AGN within a cluster and various AGN properties that relate to physical conditions within the AGN, such as obscuration. Obscured AGN generally have higher X-ray hardness ratios, lower variability significance \citep{ks07}, and steeper mid-IR power-law slopes \citep{ah06}. Figure~\ref{sigmavsdist} plots variability significance vs. distance from the cluster centre for all optical variables. We see no clear correlation between variability significance and distance from the cluster centre, though the most significant optical variables appear to lie at distances that are 20\% to 60\% of the cluster virial radius.

\begin{figure}
\centering
\includegraphics[scale=0.90]{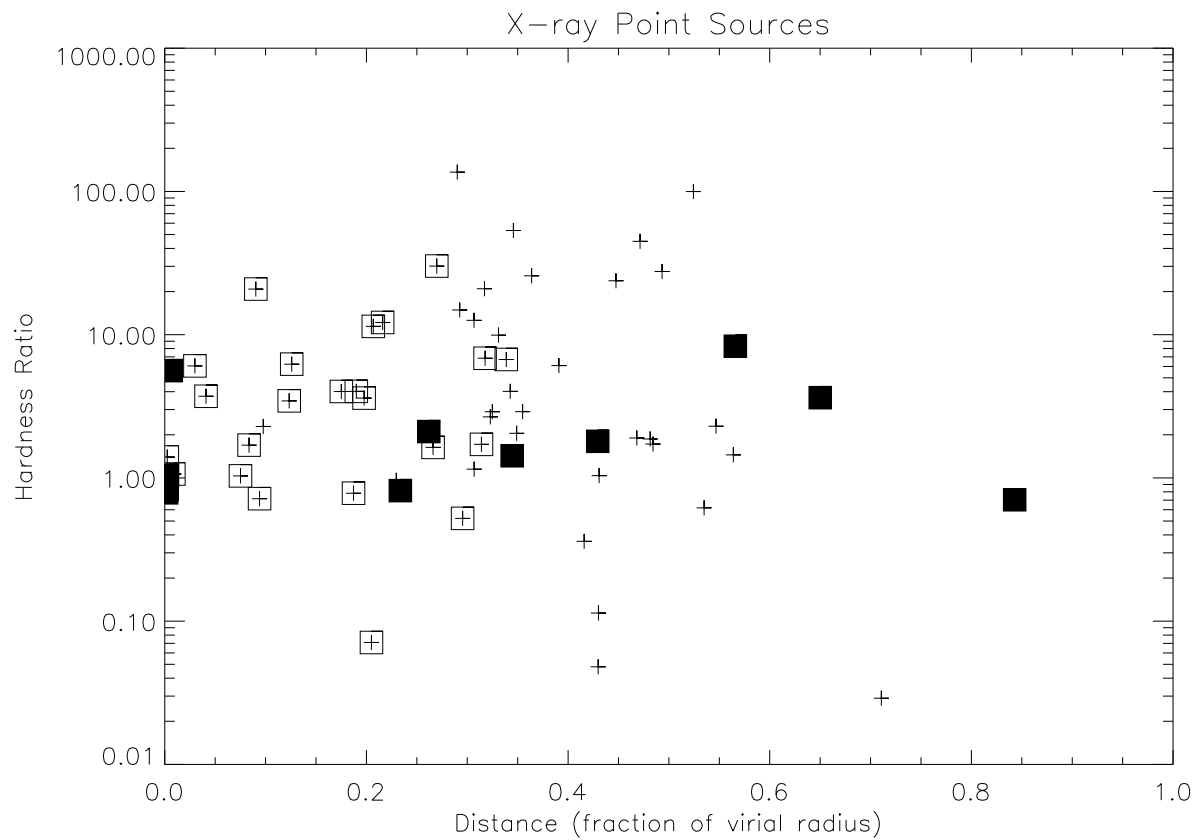}
\caption[X-ray Hardness Ratio vs. Cluster Radius]{X-ray hardness ratio vs. radial distance from the cluster centre in fraction of the virial radius. A plus symbol indicates all AGN in the cluster region, open squares indicate sources with $>$80\% cluster membership probability, and filled squares indicate spectroscopically-confirmed cluster members.}
\label{hrvsdist}
\end{figure}

Figure~\ref{hrvsdist} shows the X-ray hardness ratio vs. cluster radius for all X-ray point sources observed in both the hard and soft bands. Hardness ratio is calculated as 

\begin{equation}\label{} HR = \frac{F_{X}(2-8keV)}{F_{X}(0.5-2keV)}\end{equation}

\noindent

as done in KS12. We detect both hard and soft X-ray sources at all radii, though the hardest sources appear at radii less than $\sim$60\% of the cluster virial radius (we note that we do not detect many X-ray sources outside of this radius in general). There is no clear trend with radial distance from the cluster centre, though we find that all X-ray point sources with spectroscopically-confirmed cluster redshifts are among the softer sources and have a small range of X-ray hardness ratios, with only two cluster members having hardness ratios $>$5. This may be a selection effect, since softer X-ray sources may also be optically brighter and easier to detect in spectroscopic follow-up surveys.

\begin{figure}
\centering
\includegraphics[scale=0.90]{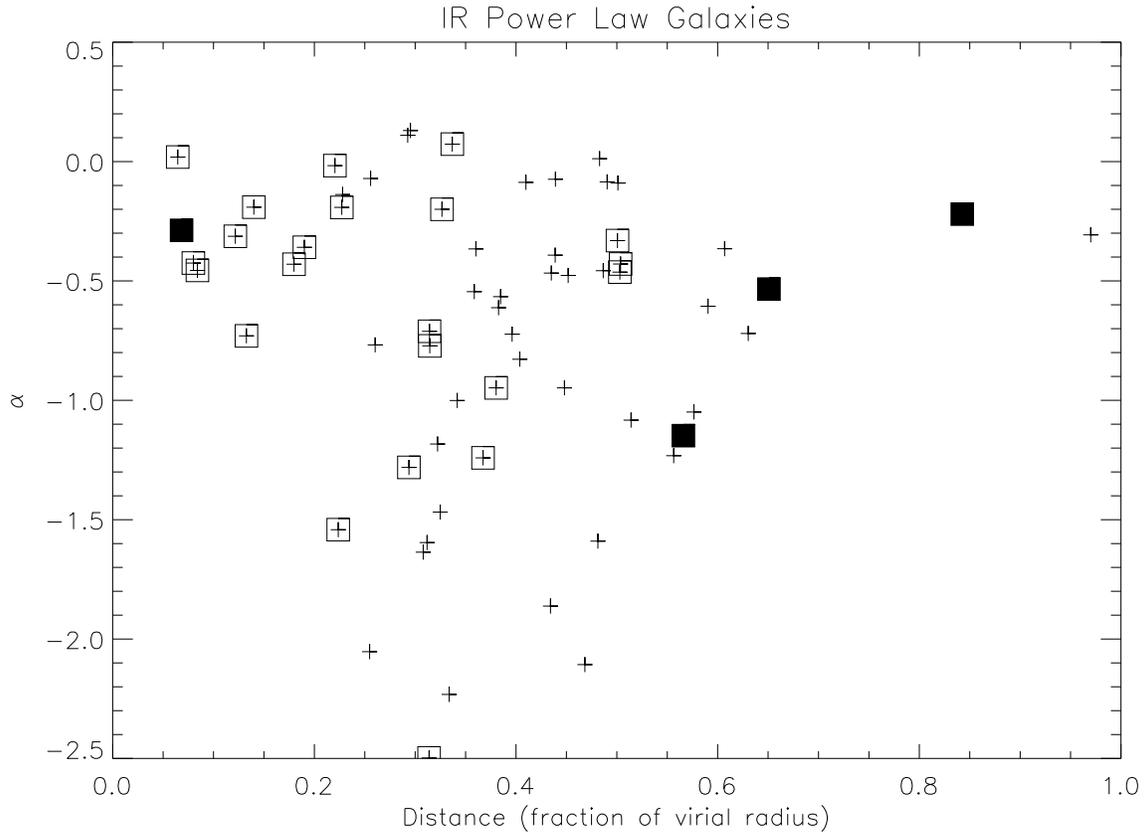}
\caption[IR Power Law Slope vs. Cluster Radius]{IR power law slope ($\alpha$) vs. radial distance from the cluster centre in fraction of the virial radius. A plus symbol indicates all AGN in the cluster region, open squares indicate sources with $>$80\% cluster membership probability, and filled squares indicate spectroscopically-confirmed cluster members.}
\label{alphavsdist}
\end{figure}

Finally, Figure~\ref{alphavsdist} plots the slope of IR power-law vs. cluster radius for all IR power-law sources. \citet{ah06} find that the slope of the power law fit relates to AGN type, where steeper (i.e., more negative) values represent NLAGNs and shallower power-law SEDs are classified as BLAGNs. We see that sources with SEDs resembling BLAGNs ($\alpha$ $>$ -0.9) are found at all radii, but NLAGN-like SEDs ($\alpha$ $<$ -0.9) appear primarily at radii between $\sim$20--50\% of the cluster virial radius. There are only four IR power-law sources with spectroscopically-confirmed cluster redshifts and only one falls between 0.2 and 0.6 of the cluster virial radius, though it does have a steeper IR SED than the other three objects.

In summary, we find no significant correlations between AGN properties such as variability significance, X-ray hardness, or IR power-law slope and the radial location of an AGN within the cluster.  

\section{AGN Host Galaxies in Clusters}
\label{colors}

The link between AGN activity and the evolution of galaxies may be reflected in the properties of galaxies which currently host actively accreting supermassive black holes. There is also a clear correlation between environment, galaxy type and star formation rate. We examine the colours of our AGN candidate host galaxies in an effort to explore the relationship between the presence of an AGN, the host galaxy, and the cluster environment.

\subsection{Variable, X-ray and IR AGN Host Galaxies}
\begin{figure}
\centering
\includegraphics[scale=0.77]{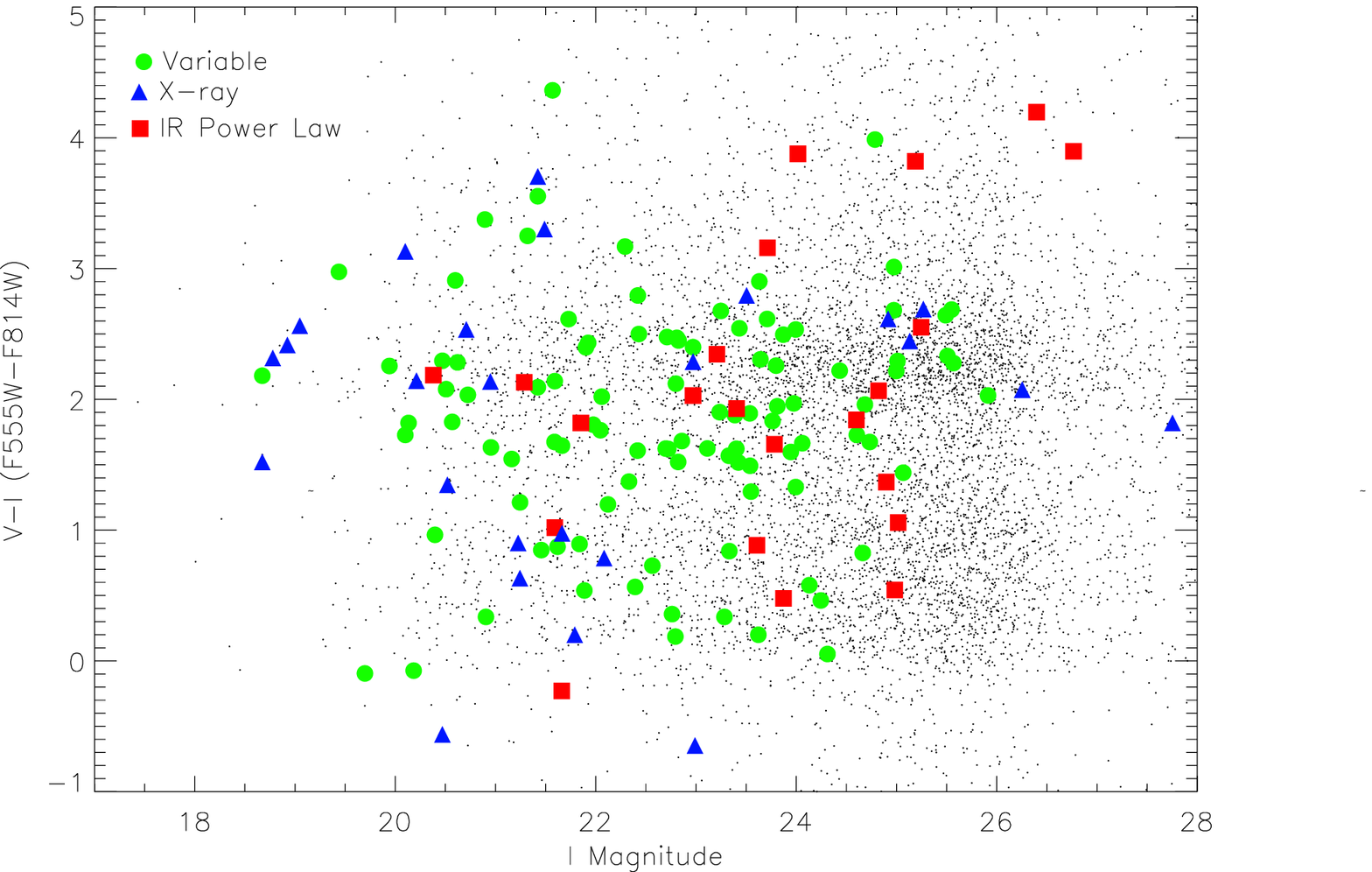}
\caption[Galaxy Colours vs. Redshift]{Colour-Magnitude diagram for all sources in the seven cluster regions at z $\sim$ 0.5. Black points are normal galaxies, green circles are optical variables, blue triangles are X-ray point sources, and red squares are IR power-law sources.}
\label{groupcolormag}
\end{figure}

As described in KS12, our cluster fields have been observed in both the HST V (F555W) and I (F814W) bands. To compare the observed V-I colours and minimize k-corrections, we limit our analysis to those clusters that fall in the small redshift range between 0.504 $<$ z $<$ 0.550, which includes seven of our 12 cluster fields. In Figure~\ref{groupcolormag}, we show the observed colour-magnitude diagram for all galaxies in the cluster field (cluster and field galaxies) for these z $\sim$ 0.5 clusters, with AGN candidates indicated by coloured symbols. The expected bimodal distribution of ``red sequence'' bulge-dominated, passively-evolving galaxies and ``blue cloud'' star-forming, disk-dominated galaxies \citep[e.g.,][]{strateva01,blanton03,kauffmann03} can be clearly seen in the normal galaxy population, indicated with black dots in the figure.

Many studies have found that the optical colours of AGN selected via X-ray, IR, and optical variability largely reflect the colours of the host galaxy. \citet{cardamone10} find that X-ray-selected AGN show less that 0.1 mag of contamination by the AGN on the host galaxy rest-frame U-V colours. \citet{hickox09} find that only 0.4--0.5 mag of correction in u-r is required to remove AGN contamination from the bluest X-ray and mid-IR sources in their field AGN survey. We expect the most luminous blue AGN candidates to be those with the most colour contamination from the nuclear emission on the host galaxy light. An AGN-dominated galaxy would lie in the blue cloud at the bright end of the distribution in Figure~\ref{groupcolormag}. We find only five AGN candidates among blue sources at I $<$ 20 (corresponding to an absolute magnitude of M$_{I}$ $\sim$ -22 at a redshift of 0.5 and M$_{I}$ $\sim$ -23.5 at a redshift of 0.9). Thus, we do not see evidence for a large amount of AGN contamination among the host galaxies in our AGN sample and assume that the colours represent the AGN host galaxy light as a reasonable approximation. We note, however, that some fraction of AGN sources may be shifted towards bluer colours. We estimate this effect to be less than half a magnitude in the most extreme cases.

\begin{figure}
\centering
\includegraphics[scale=0.77]{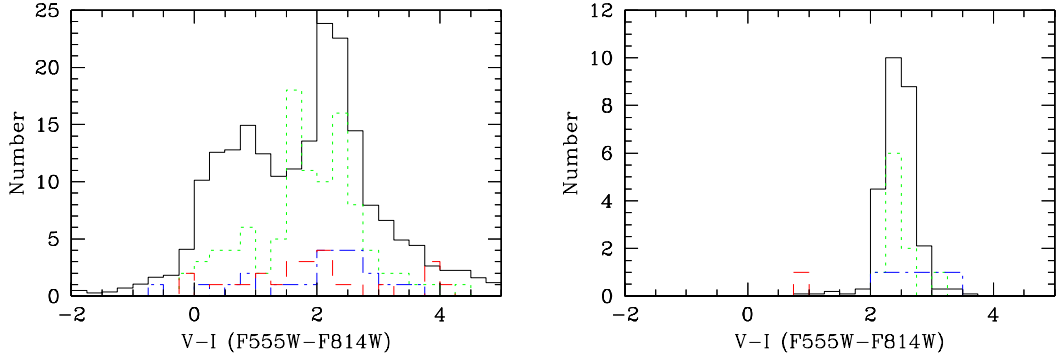}
\caption[Host Galaxy Colour Distribution]{Histogram of observed V-I host galaxy colours for all sources in the seven cluster regions at z $\sim$ 0.5. The left panel includes all sources within the cluster regions and the right panel includes spectroscopically-confirmed cluster members only. The black (solid line) histogram represents normal galaxies, green (dotted line) is optical variables, red (dashed line) is IR power law sources, and blue (dot-dashed line) is X-ray point sources. The galaxy histogram is divided by 35 and the AGN histograms are divided by 10 for comparison purposes.}
\label{groupcolors}
\end{figure}

Figure~\ref{groupcolors} shows the distribution of observed V-I colour for both normal galaxies and those hosting AGN for our z $\sim$ 0.5 clusters. The left panel includes all sources in the cluster field (cluster and field galaxies/AGN) and the right panel contains only spectroscopically-confirmed cluster members. The spectroscopically-confirmed members are mainly red galaxies with V-I $>$ 2. This reflects the selection effect for the spectroscopic sample, since most objects for which spectra were obtained were chosen based on their red colours \citep{barrett06}. For this reason, the spectroscopically-confirmed cluster member colour distribution is not a good measure of the true galaxy distribution in our clusters. We rely on the distribution of all sources in the cluster fields and compare to pure field survey results to determine if a difference can be detected.

The left panel containing all sources in the cluster field reveals optical variables extending over a broad colour range, peaking in the ``green valley'' between the red sequence and blue cloud at V-I $\sim$ 1.5 with an average colour of $\langle$V-I$\rangle$ $\sim$ 1.9. There are about 100 optically variable AGN in our z $\sim$ 0.5 cluster fields, compared to only $\sim$25 IR and X-ray-selected AGN. Even with this small number of AGN, we see that the IR-selected and X-ray-detected AGN follow a similar trend, with a range of colours and $\langle$V-I$\rangle$ = 2, though there is no obvious peak in either distribution. 

\citet{hickox09} examined the host colours of X-ray- and IR colour-selected AGN at 0.25 $<$ z $<$ 0.8. They found that X-ray AGN tend to lie in the ``green valley'' with a small tail of blue host galaxies. The distribution of their IR AGN sample is similar to both ours and that of \citet{atlee11}, which show IR AGN residing in galaxies that have slightly bluer hosts than the X-ray AGN and a less pronounced green peak. We also compare our results to \citet{sarajedini11}, who examined the colours of GOODS field AGN identified via optical variability, X-ray emission, and IR power-law SED fits. They found that X-ray sources and optical variables inhabit host galaxies with a range of colours, with the X-ray point sources peaking in the green valley and the variables showing a flatter distribution through the green valley. Their IR power-law sources showed a relatively flat distribution across the range of U-V colours. These results agree with our findings as well, indicating that the host galaxy colours of AGN in these cluster regions do not differ significantly from that of pure field galaxy surveys. Since this conclusion is based on a sample that contains a mixture of cluster and field galaxies/AGN, the cluster environment cannot be ruled out as having no effect on AGN host galaxy colour because the impact may be too subtle to discern in our mixed population.

\citet{hickox09} present a general picture of galaxy evolution in accordance with the model presented by \citet{hopkins08}, where galaxies begin as disk-dominated gas-rich, star-forming systems characterised by blue optical colours. Once a galaxy undergoes some sort of triggering event, such as a major merger, it displays a relatively short ($\sim$10$^{8}$ yr) phase of AGN activity, which is subsequently quenched along with the star formation in the galaxy. As AGN and star formation activity decline and disappear, the host galaxy's spectrum evolves toward intermediate colours that result from a composite spectrum of emission from the older stellar population and a declining contribution from younger stars. Finally, after $\sim$1--2 Gyr, the galaxy evolves into a red, passively-evolving spheroid-dominated system in the red sequence portion of the colour-magnitude diagram. Several studies have shown that AGN hosts have colours that indicate this decline in star formation and transition from blue to red \citep{schawinski07,silverman08,bundy08}. These findings support the idea that nuclear activity suppresses star formation through feedback mechanisms such as radiation pressure and/or jets, which can inject energy (either radiative or mechanical) into the gas surrounding the nucleus and prevent it from cooling and fuelling any further star formation \citep[e.g.,][]{fabian03,mcnamara05,forman07,mn07}. 

The dense cluster environment is expected to have some effect on the star formation rates and colours of cluster galaxies. Studies have found evidence of star formation being triggered in infalling galaxies, which later couples with physical mechanisms such as ram pressure stripping or galaxy harassment to quench star formation \citep{bai07,marcillac07}. \citet{chung10} found evidence for an elevated IR luminosity function in the Bullet Cluster, a galaxy cluster at z = 0.296 that is currently undergoing a major supersonic merger. The observed excess in star-forming IR galaxies can be associated with the infalling group of galaxies that have not yet been processed by the cluster environment into more quiescent galaxies. This suggests that the processes responsible for quenching star formation occur on a longer timescale than that of the infalling group's accretion into the Bullet Cluster ($\sim$250 Myr). One such slower process, known as ``strangulation,'' can occur in galaxies being assimilated into galaxy clusters. A galaxy's loosely-bound halo gas is gently pushed away through interactions with the intracluster medium, which can affect both the morphology and star formation rate of the galaxy \citep{larson80,balogh00}. This longer timescale for cluster environment effects on galaxy colours may be the reason we do not see an obvious difference between the AGN hosts in clusters when compared to field galaxy studies. The feedback mechanism that may affect the AGN hosts individually could also be operating on a larger scale in the ICM. Relativistic jets from radio AGN (such as the central cD galaxy in a cluster) could heat the surrounding intracluster medium and aid in quenching star formation in cluster galaxies \citep[e.g.,][]{birzan04,rafferty06,croston08}. To further investigate the cluster environmental effects and possible feedback effects within individual AGN, we compare the AGN host colours to the normal galaxies as a function of cluster radius in Section~\ref{radcolors}.

\subsection{Cluster Morphology and AGN Host Galaxies}

\begin{figure}
\centering
\includegraphics[scale=0.90]{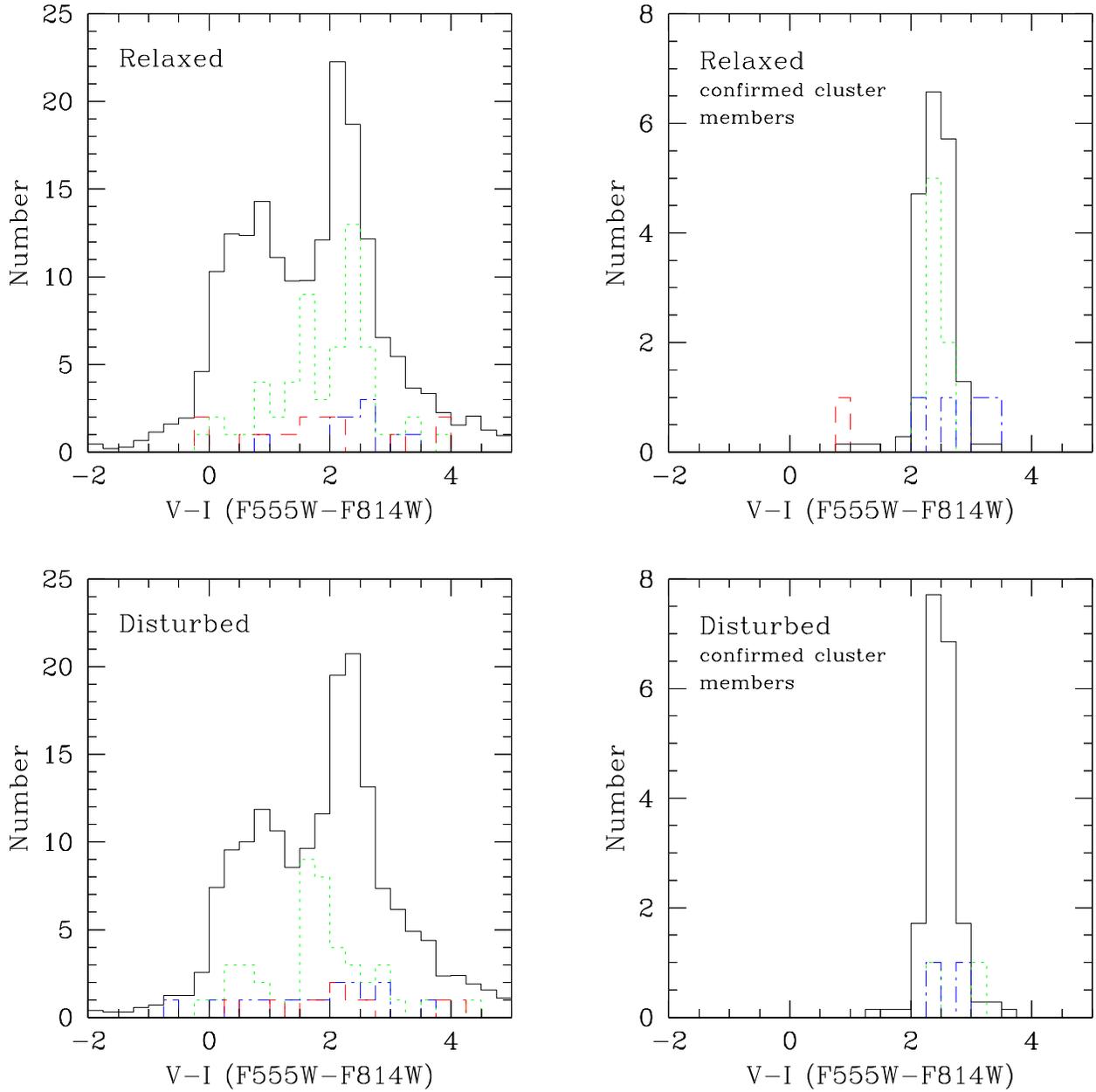}
\caption[Galaxy Colours vs. Cluster Morphology]{Histogram of observed V-I galaxy colours in relaxed (top panels) and disturbed clusters (bottom panels). Three of our clusters at z $\sim$ 0.5 are classified as disturbed, while four clusters at z $\sim$ 0.5 are classified as relaxed. The left panels include all sources found in the cluster regions and the right panels include only spectroscopically-confirmed cluster members. The black (solid line) histogram is normal galaxies, green (dotted line) is optical variables, red (dashed line) is IR power law sources, and blue (dot-dashed line) is X-ray point sources. The galaxy histogram is divided by 20 and the AGN histograms are divided by 7 for comparison purposes.}
\label{morphcolors}
\end{figure}

We investigate the galaxy colours in clusters divided by morphology classification to further explore the interplay of AGN and cluster environment on AGN host galaxies. Figure~\ref{morphcolors} shows a histogram of the optical V-I colours of the galaxies and AGN in relaxed (top panels) and disturbed clusters (bottom panels). The figures include data from our seven z $\sim$ 0.5 cluster fields, which are fairly evenly divided among disturbed (3) and relaxed (4) clusters.

The left panels of the figure include all galaxies in the cluster fields (i.e., field plus cluster sources). The disturbed clusters reveal a distribution of optical variables peaking in the green valley at V-I = 1.65 with $\langle$V-I$\rangle$ = 1.9. In relaxed clusters, the variables have roughly the same mean V-I colour, but extend through the green valley and peak in the red sequence at V-I = 2.3. While the colour distributions are not significantly different (we calculate a KS probability of 82\% that they are drawn from different parent populations), the absence of the red peak among the disturbed clusters is noticeable. The X-ray and IR AGN have host galaxy colours distributed throughout the full range of colours in both relaxed and disturbed clusters, with similar $\langle$V-I$\rangle$ $\sim$ 2. Among the spectroscopically-confirmed cluster members (right panels), the colour selection effect is apparent. The distribution here mainly reflects that of the left panels limited to sources redder than V-I = 2, where the spectroscopic follow-up observations were targeted.

If galaxy interactions and substructure in the hot X-ray gas are more common in disturbed clusters than relaxed clusters, this may be the reason that the AGN hosts appear to tend toward bluer colours in the disturbed clusters. These clusters are characterised by X-ray substructure and at least two of these clusters are either currently undergoing a merger or have recently experienced one in the past. Minor mergers, galaxy harassment, and interactions with substructure in the intracluster medium in disturbed clusters may be more common than in the X-ray-smooth, structurally-symmetric relaxed clusters. These processes may be responsible for triggering or maintaining star formation or AGN activity in cluster galaxies, pushing them toward bluer colours rather than passively-evolving red sequence galaxies. In contrast, galaxies in relaxed clusters may undergo fewer interactions as they fall into the smoother cluster density profile, and thus may be evolving more passively than those in disturbed clusters.

\section{Radial Distribution of AGN Host Galaxy Colours}
\label{radcolors}

In the dense cluster environment, star formation activity in galaxies is subdued compared to field galaxies and a higher fraction of early-type galaxies is observed \citep[e.g.,][]{hubble36,dressler80}. Furthermore, there is a well-known morphology-density relation observed in clusters, which is characterised by a higher fraction of early-type galaxies in dense cluster cores and an increasing fraction of late-type galaxies with increasing distance from the cluster centre as densities approach that of the field \citep[e.g.,][]{dressler80,oemler74,ms77}. A galaxy's distance from the centre of a cluster may also correlate with the time since infall into the cluster \citep{gao04}, as galaxies at large cluster radii have not yet encountered the densest parts of the cluster while galaxies in the cluster core were either formed in the dense environment or have already crossed from the outskirts to the cluster core at least once. 

\begin{figure}
\centering
  \begin{tabular}{@{}cccc@{}}
    \includegraphics[scale=0.45]{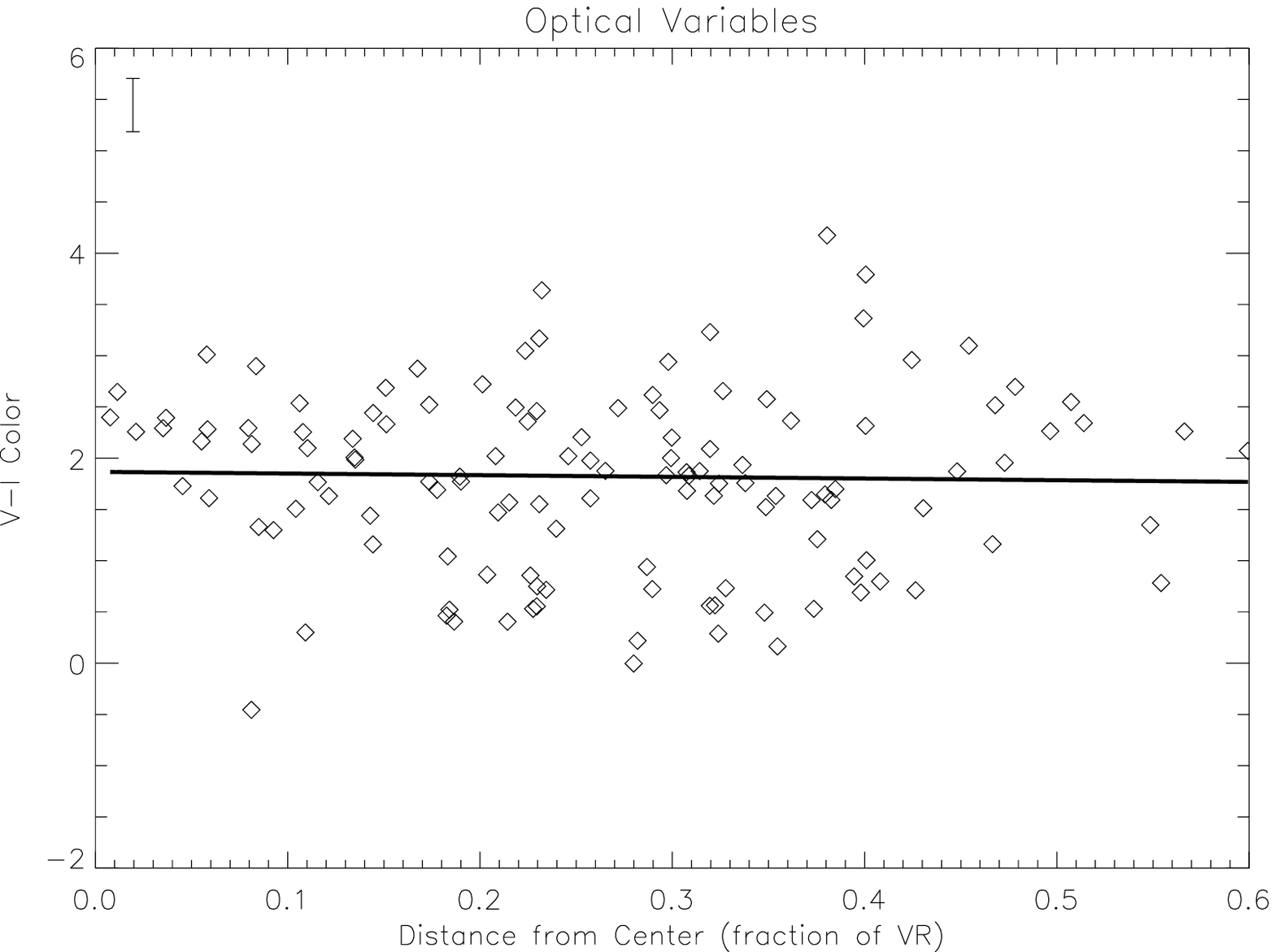} &
    \includegraphics[scale=0.45]{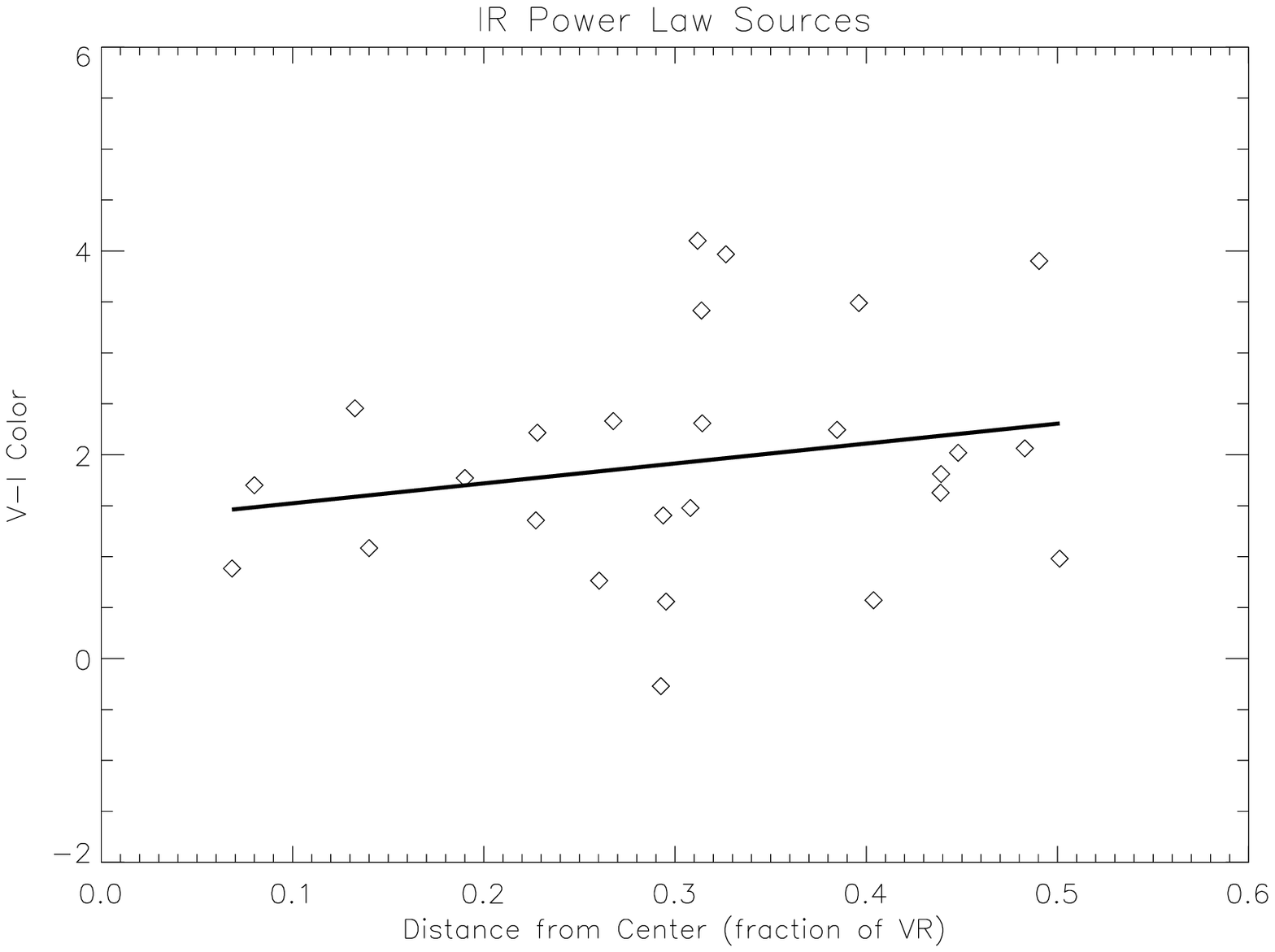} \\
    \includegraphics[scale=0.45]{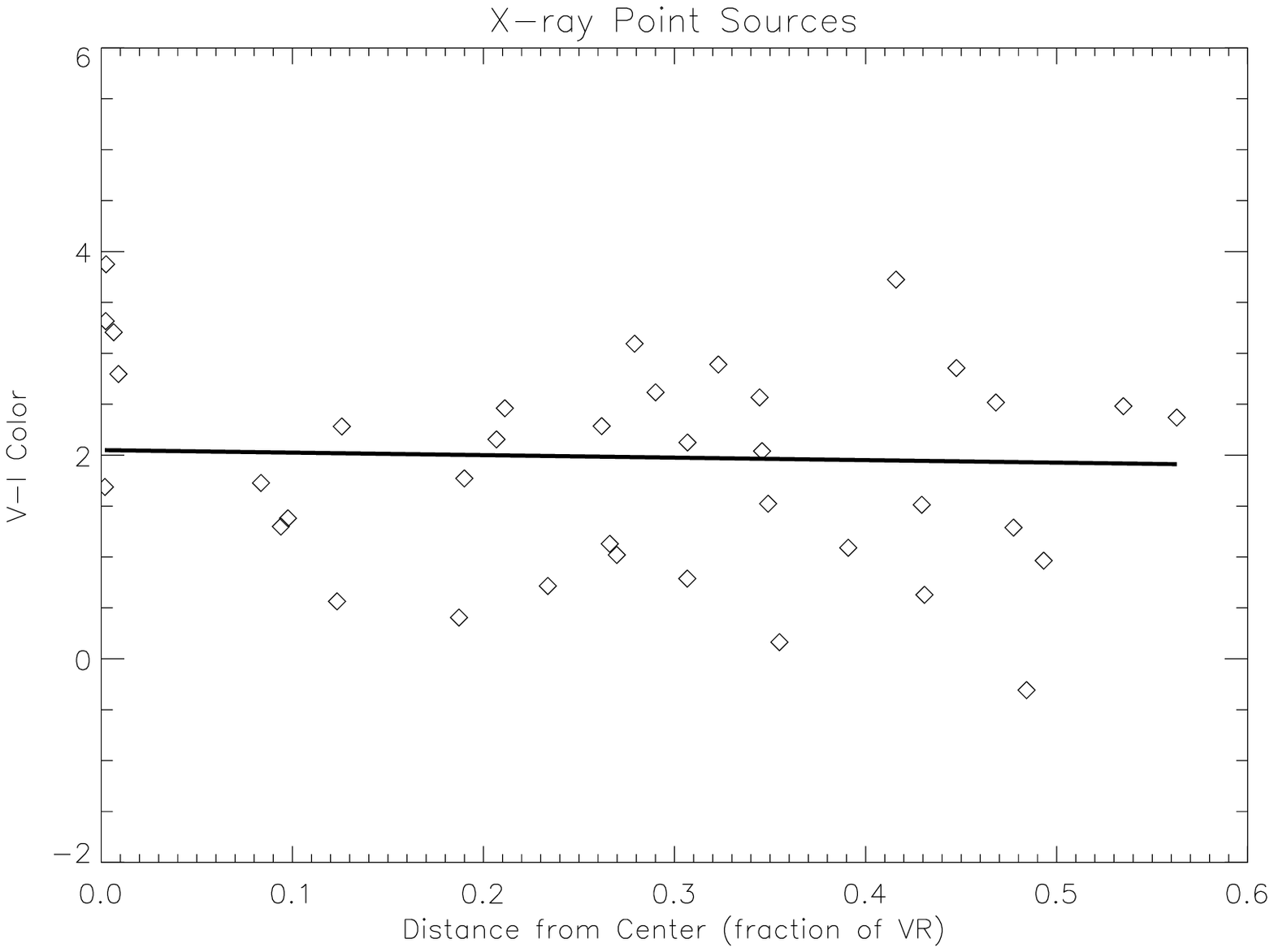} &
    \includegraphics[scale=0.45]{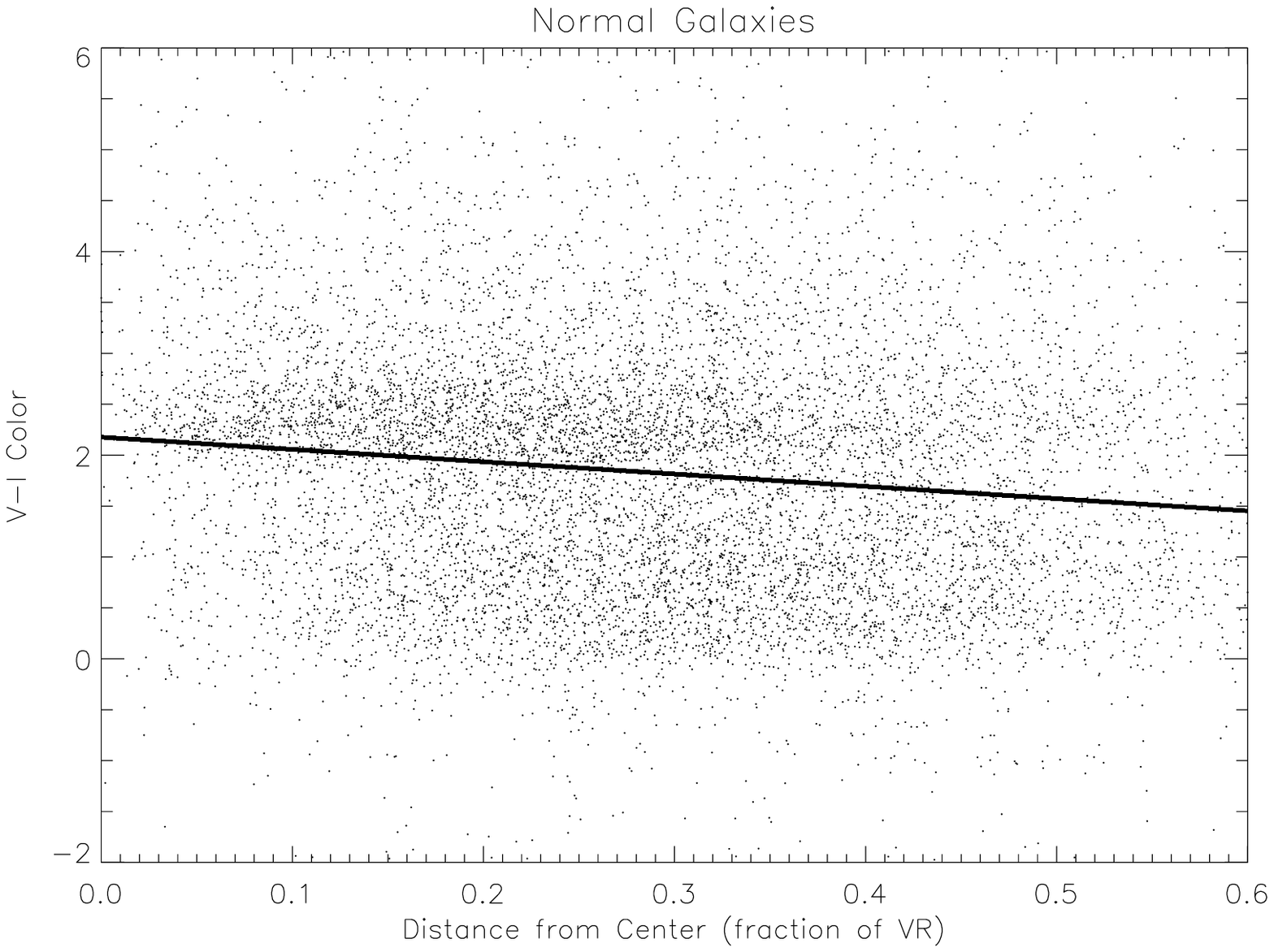}   \\
  \end{tabular}
  \caption[Galaxy Colours By Type vs. Cluster Radius]{V-I colour vs. distance from the cluster centre as a fraction of the virial radius for all galaxies and AGN host galaxies in the nine cluster fields at z $<$ 0.7. The solid line is a linear fit to the points. {\it Top Left}: Optical Variables; {\it Top Right}: IR Power Law Sources; {\it Bottom Left}: X-ray Point Sources; {\it Bottom Right}: Normal Galaxies. The error bar in the upper lefthand corner of the top left plot indicates the average photometric error in all cases.}
\label{slopesbytype}
\end{figure}

Given that the presence of an AGN may also affect star formation in the host galaxy and thus impact the galaxy colour, we compare the radial colour distribution of the host galaxies of our AGN candidates with that of the normal galaxy population to determine whether the AGN has any significant effect on the host galaxy as a function of radial distance from the cluster centre. In Figure~\ref{slopesbytype}, we show the radial colour distribution of the host galaxies for our three types of AGN candidates and the normal galaxy population for all galaxies within 60\% of the virial radius of the cluster. These figures include all sources within the cluster region and therefore contain both cluster and field galaxies/AGN. We have combined the data for the nine cluster regions at redshifts between z $\sim$ 0.5 and z $\sim$ 0.7 to provide the largest number of sources while avoiding significant redshift effects. We conducted this analysis using all sources in the cluster region, even though the sample will be contaminated by foreground and background galaxies. Since the spectroscopically-confirmed sources have a severe colour selection effect, any trend with radius has already been effectively removed from that sample. Likewise, our sources with greater than 80\% cluster membership probability have also had a colour and radial selection criterion imposed upon them, which would impact our analysis. Therefore, only the sample that includes all sources in the cluster fields is free from imposed selection effects. 

To identify radial colour trends, the distributions were fitted with a straight line using $\chi^{2}$ minimisation to determine the best fit and the error on the fit. The slopes of the fits to the optical variables, IR power-law sources, and X-ray sources are -0.16 $\pm$ 0.6, 1.95 $\pm$ 1.8 and -0.25 $\pm$ 1.2, respectively. Combining the AGN samples yields a slope of -0.07 $\pm$ 0.54, consistent with a slope of zero. The normal galaxy distribution colour slope was found to be -1.2 $\pm$ 0.10. Thus, the normal galaxies are generally bluer with increasing radius and demonstrate a significantly steeper relation than the host galaxies of any AGN type or the combined AGN sample, which show no significant relation between cluster radius and colour. We found that cluster morphology (relaxed vs. disturbed) has no discernible effect on the slopes of the fits to the AGN or the normal galaxy distribution. 

The galaxies and AGN analyzed with these fits represent the cluster plus field populations, where the field galaxies are expected to have no colour relation with cluster radius. Thus, any observed slope of this line is partly due to the true colour gradient expected among cluster galaxies and partly due to the decreasing cluster density and increasing dominance of the bluer field population as a function of increasing cluster radius. Since our fit to the AGN host galaxies also includes all sources in the cluster field, the AGN samples are subject to the same changes in cluster density and field contamination as a function of radius. The negative slope (with $>$3$\sigma$ confidence) of the normal galaxies reveals that the cluster population is dominant enough to reveal this expected trend. As discussed previously, several studies have found a relationship between normal galaxy colour and cluster radius. This relationship has been interpreted as due to the slow quenching of star formation over timescales of a few Gyr \citet{vdl10}, possibly through the process of strangulation (\citet{chung10,balogh00}). Our observations of the normal cluster galaxy population are consistent with these results. 

The lack of a similar relationship among the AGN host galaxy colours is interesting, since the AGN population should contain a similar mix of cluster and field sources as function of cluster radius. While the colour of a normal galaxy within a cluster appears to depend to some extent on the radial distance and consequently the local environment, the AGN host colours do not reveal this dependency. This observation supports the possibility that processes related to the accreting supermassive black hole have a more significant impact on the star-forming properties of the host galaxy than does the intracluster medium and/or the local environmental density.

\section{Conclusions}
\label{conclusions}

We have explored several issues concerning the AGN population in dense cluster environments by analysing 12 galaxy clusters at redshifts 0.5 $<$ z $<$ 0.9 to determine the AGN fraction in clusters and address issues of AGN triggering and fuelling in massive galaxy clusters. In our first paper, \citep{ks12a}, we compiled a catalog of cluster AGN candidates using a combination of three detection techniques: optical variability, X-ray point source detection, and mid-IR power-law SEDs. We also discussed the overall AGN fraction in our cluster sample. In this paper, we focused on the analysis of the radial distributions of AGN and colours of AGN host galaxies in clusters.

We found that the radial distribution of the AGN population is consistent with the radial distribution of normal galaxies in two-thirds of our cluster fields, while one-third reveal an AGN population distributed differently than the normal galaxies. Among those with significantly different distributions, the AGN are generally less centrally concentrated than the normal galaxies except within the central 20\% of the virial radius, where the AGN are either more centrally concentrated or follow the normal galaxy distribution closely. When we combine the distributions of our cluster regions scaled by virial radius, we find that the AGN distribution rises slightly faster than the normal galaxies in the inner 20\% of the clusters, mainly due to the X-ray-selected AGN, and then becomes less centrally concentrated than the normal galaxies at greater radii. The slight over-concentration of AGN toward the cluster centres may be explained by interactions (e.g., mergers, tidal interactions) with the cD galaxies and other giant ellipticals in the cores of these clusters. At intermediate cluster radii, the cumulative distribution of AGN appears to rise more slowly than the normal galaxies and may be explained by the following scenario: as galaxies fall into the cluster for the first time, nuclear activity may be triggered by interactions at the cluster-field interface. The AGN triggered as a result will eventually deplete their accretion disks over a timescale less than the cluster crossing time, thus resulting in a less centrally concentrated AGN distribution at intermediate radii. Since our observations do not extend to the virial radius, we cannot confirm a higher concentration of AGN at the cluster virial radius.

We find that relaxed clusters, those with a clear centre and smooth radial distribution of galaxies, have more centrally-concentrated AGN than the normal galaxy distribution at $>$95\% confidence and these are primarily X-ray selected AGN. The disturbed clusters have AGN that are less centrally concentrated than the normal galaxies. In both types of clusters, the IR power-law sources appear to be the least centrally concentrated AGN type. Part of the reason for the difference between relaxed and disturbed clusters may be due to the fact that disturbed cluster radial distributions are spread over a larger range of radii since they do not have well-defined cores or spherical symmetry. This could make the detection of central concentrations difficult to identify in this type of cluster. 

We explored correlations between the radial location of galaxies hosting AGN within the cluster field and various AGN properties that relate to physical conditions within the AGN. However, we found no significant correlations between the optical variability significance, X-ray hardness ratio, or IR power-law slope and the radial location of the AGN candidates within the cluster fields.  

The host galaxies of the AGN candidates in our survey appear to have a different colour distribution than the normal galaxies in our cluster fields. The AGN tend to occupy the ``green valley" avoided by most normal galaxies. Comparing our results with the GOODS field, we found that the host galaxy colours of AGN in the cluster fields are not significantly different from pure field samples. While we cannot rule out the impact of cluster environment effects on these distributions, we note that the effect may be subtle and thus consistent with the quenching of star formation taking place over longer timescales, as found in recent cluster studies. 

In disturbed clusters, the distribution of AGN candidates peaks in the green valley, whereas in relaxed clusters the distribution extends through the green valley but peaks among redder galaxies, especially among the optically variable AGN candidates. If galaxy interactions or substructure in the hot X-ray gas is more common in disturbed clusters than relaxed clusters, minor mergers, galaxy harassment, and interactions with substructure in the intracluster medium may be more common as a result of this cluster morphology, which is not present in the structurally-symmetric relaxed clusters. This may be responsible for triggering more star formation in the AGN hosts within disturbed clusters, pushing the galaxies toward bluer colours instead of allowing their star formation to be mostly quenched and the hosts to transition to passively-evolving red sequence galaxies.

Given that the presence of an AGN as well as the intracluster medium may affect star formation in the host galaxy and thus impact the galaxy colour, we have examined the radial colour distribution of the host galaxies of our AGN candidates compared with that of the normal galaxy population in our cluster fields. In this way, we attempted to determine if the presence of an AGN has any significant effect on the host galaxy as a function of radial distance from the cluster centre. We found that the AGN hosts display no apparent change in colour with cluster radius, while the normal galaxies become bluer at greater radii with $>$3$\sigma$ confidence. The fact that we see no colour-radius relation among AGN hosts supports the theory that processes related to the accreting supermassive black hole have a more significant impact on the star-forming properties of the host galaxy than does the intracluster medium and/or the local environmental density.

\section{Acknowledgements}
\label{acknowledgements}
The results in this paper are based on observations made with the NASA/ESA Hubble Space Telescope, obtained at the Space Telescope Science Institute, which is operated by the Association of Universities for Research in Astronomy, Inc., under NASA contract NAS 5-26555. These observations are associated with program \# GO 10493, cycle 14 and GO 10793, cycle 15. This work is based in part on archival data obtained with the Spitzer Space Telescope, which is operated by the Jet Propulsion Laboratory, California Institute of Technology under a contract with NASA. This research has made use of data obtained from the Chandra Data Archive and the Chandra Source Catalog, and software provided by the Chandra X-ray Center (CXC) in the application packages CIAO, ChIPS, and Sherpa (ObsIDs 512, 520, 529, 902, 913, 1654, 1655, 1656, 1657, 3180, 3196, 3197, 3199, 3529, 3581, 3584, 3585, 3587, 3589, 3595, 4195, 4200, 5011, 5012, 5014, 5794). Funding was provided by NSF CAREER grant 0346691. We also thank the referee for a careful reading of this work and helpful comments and suggestions, which improved the quality of this paper.

\bibliography{references}

\clearpage
\newpage

\label{lastpage}

\end{document}